\documentclass[aps,showpacs,prl,twocolumn,final]{revtex4}

\usepackage{graphicx}% Include figure files
\usepackage{dcolumn}% Align table columns on decimal point
\usepackage{bm}% bold math
\usepackage{color}
\usepackage{amsmath,amssymb,mathrsfs,array,longtable,delarray}

\begin{document}

\title{On the Zero-Bias Anomaly in Quantum Wires}

\author{S. Sarkozy}
\altaffiliation{On leave from Northrop Grumman Space Technology, One Space Park,
Redondo Beach, California, 90278} \altaffiliation{e-mail: stephen.sarkozy@ngc.com}
\author{F. Sfigakis}
\altaffiliation{e-mail: fs228@cam.ac.uk}
\author{K. Das Gupta}
\author{I. Farrer}
\author{D. A. Ritchie}
\author{G. A. C. Jones}
\author{M. Pepper}
\affiliation{Cavendish Laboratory, J. J. Thompson Avenue, Cambridge, CB3 OHE,
United Kingdom}
%\date{\today~---~version 2.02}

\begin{abstract}
Undoped GaAs/AlGaAs heterostructures have been used to fabricate quantum wires in
which the average impurity separation is greater than the device size. We compare
the behavior of the Zero-Bias Anomaly against predictions from Kondo and spin
polarization models. Both theories display shortcomings, the most dramatic of which
are the linear electron-density dependence of the Zero-Bias Anomaly spin-splitting
at fixed magnetic field $B$ and the suppression of the Zeeman effect at pinch-off.
\end{abstract}


\maketitle

Split gates \cite{Thornton86} can be used to restrict transport from a
two-dimensional electron gas (2DEG) to a ballistic one-dimensional (1D) channel.
This results in the quantization of the differential conductance
$G=dI/dV_{\text{sd}}$ in units of $G_0=2e^2/h$ at zero magnetic field
\cite{vanWees88etal,Wharam88etal}. A shoulder on the riser of the first quantized
plateau, the ``0.7 anomaly'' or ``0.7 structure'' \cite{Thomas96etal}, is not
completely understood but generally acknowledged to result from electron-electron
interactions. Although spin polarization models
\cite{Wang96,Kristensen00etal,Reilly02etal,Abi07,Francois08Betal,Berggren08} and 1D
Kondo physics models \cite{Cronenwett02etal,Meir02,Rejec06} can describe many
experiments, neither can explain all phenomena associated with the 0.7 structure.
One example is the so-called zero-bias anomaly (ZBA): a peak in $G$ centered at
$V_{\rm{sd}}=0$ for $G\!<\!2e^2/h$ when sweeping source-drain bias $V_{\text{sd}}$
at a fixed gate voltage $V_{\text{\tiny{gate}}}$ at low temperature $T$. Spin
polarization models cannot alone predict its occurrence in quantum wires, although
an embedded impurity near or in the 1D channel could produce a ZBA via the 0D Kondo
effect \cite{Meir93,Goldhaber98Aetal,Cronenwett98etal,vanWiel00etal}. On the other
hand, in 1D Kondo physics models, a bound state forms when $G < G_0$. In this
context, a resonance observed by a non-invasive detector capacitively coupled to a
quantum wire at threshold \cite{Yoon07etal} as well as a triple-peaked structure in
$G$ at fixed $V_{\text{\tiny{gate}}}$ below the 0.7 structure
\cite{Francois08Aetal} are consistent with the presence of a localized state in 1D
channels.

Systematically studying the ZBA in modulation-doped 2DEGs has proven difficult
because of the large variability of its characteristics from device to device
\cite{AbiPRBsubmit,JonGnopaper}, probably due to the randomly fluctuating
background potential caused by the ionized dopants, significant even with the use
of large ($\geq$75 nm) spacer layers. This disorder is so pervasive that one can be
led to wonder whether the ZBA always results from interactions between conduction
electrons and a random localized state near the 1D channel. However, disorder can
be dramatically reduced in undoped GaAs/AlGaAs heterostructures where an external
electric field (via a voltage $V_{\text{\tiny{top}}}$ on a metal top gate)
electrostatically induces the 2DEG \cite{Harrell99etal,SarkozyAPLsubmit}. Figure
1(a) shows the advantages of this technique, particularly at low carrier densities
(see also Fig.~3 in Ref.~\cite{Harrell99etal}), a regime most relevant for the ZBA.

In this Letter, we report on the study of the ZBA in ten quantum wires fabricated
in undoped GaAs/AlGaAs heterostructures. We demonstrate that an unsplit ZBA does
not result from interactions between conduction electrons and a random localized
state near the 1D channel: it is a fundamental property of 1D channels, in
disagreement with spin polarization models. Another inconsistency is a
suppression of the Zeeman effect at pinch-off. In disagreement with Kondo theory, we
observe a non-monotonic increase of the Kondo temperature $T_{\text{\tiny{K}}}$
with $V_{\text{\tiny{gate}}}$, and a linear peak-splitting of the ZBA with
$V_{\text{\tiny{gate}}}$ at a fixed $B$.

\begin{figure}
    \includegraphics[width=\columnwidth]{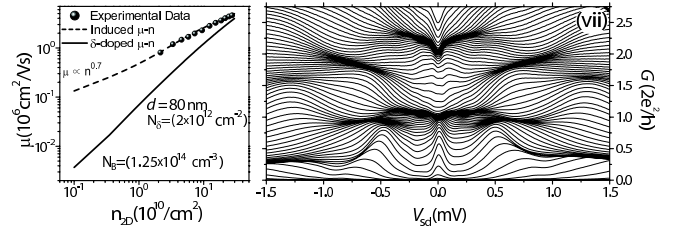}
        \caption{(a) Measured (spheres) and calculated (dashed line)
         $\mu-n_{\text{\tiny{2D}}}$ relation for T622. For comparison, we simulate an
         otherwise identical 2DEG with a $\delta$-doped layer 80 nm above. (b) $G$
         vs.~$V_{\text{sd}}$ incrementing $V_{\text{\tiny{gate}}}$ (in steps of 0.3 mV)
         of a quantum wire in an undoped heterostructure ($T=60$ mK). A ZBA can be
         observed in the riser of the $2e^2/h$ plateau.}
         \label{Fig1}
\end{figure}

The two wafers primarily used in this study, T622 (T623) with a 317 (117) nm deep
2DEG, were grown by molecular beam epitaxy and consisted of: a 17\,nm GaAs cap, 300
(100) nm of Al$_{.33}$Ga$_{.67}$As/GaAs, 1\,$\mu$m of GaAs, and a 1\,$\mu$m
superlattice with a 5\,nm Al$_{.33}$Ga$_{.67}$As/5\,nm GaAs period. No layer was
intentionally doped. For T622,
$n_{\text{\tiny{2D}}}=(0.275\,V_{\text{\tiny{top}}}/\text{V}-0.315)\times10^{11}$
cm$^{-2}$. Figure 1(a) shows the mobility $\mu$ versus the 2D sheet carrier density
$n_{\text{\tiny{2D}}}$ for T622; wafer T623 has slightly higher mobilities, e.g.
1.7$\times 10^6$\,cm$^{2}$/Vs versus 1.6$\times 10^6$\,cm$^{2}$/Vs at
5$\times$10$^{10}$\,cm$^{-2}$.  Using Matthiessen's rule far from the localization
regime, the experimental data is fit to standard models of scattering times
$\frac{1}{\tau_{total}}=\sum_j\frac{1}{\tau_j}$ \cite{Ando82-A,Gold88}. The
dominant sources of scattering in our system (analyzed in
Ref.~\cite{SarkozyAPLsubmit}) are charged background impurities and interface
roughness, from which we extracted the background impurity concentration
$N_B\!=\!1.25\times10^{14}$\,cm$^{-3}$. Intersecting the background impurity
potential with a 2DEG wavefunction of width $\lambda\!\leq\! 20$\,nm yields a
minimum average distance between scattering centers $D=0.6\,\mu$m in wafer T622. A
similar number is found for wafer T623.

\begin{figure}
    \includegraphics[width=\columnwidth]{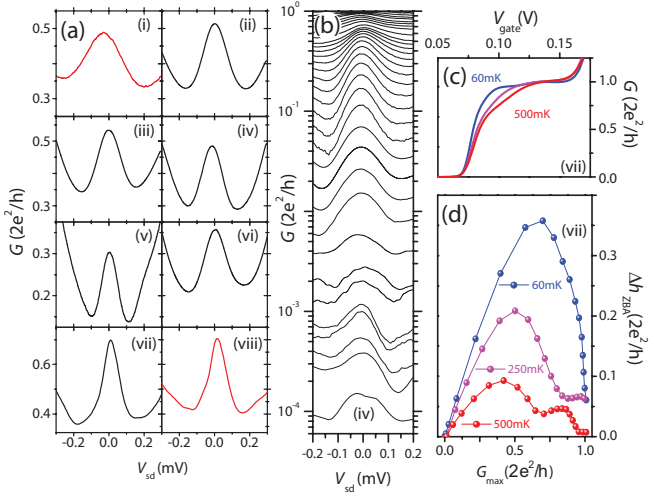}
        \caption{(color online) (a) $G$ vs.~$V_{\text{sd}}$ incrementing
        $V_{\text{\tiny{gate}}}$ for eight quantum wires, labeled (i)
        through (viii). (b) For a wide range of $G$
        (on a log scale), the ZBA occurs far beyond the ballistic
        regime ($T\!\approx\!150$\,mK). (c) $T$ dependence of the 0.7 structure at
        $V_{\text{sd}}\!=\!0$.
        (d) $\Delta h_{\text{\tiny{ZBA}}}$
        (defined in main text) for various $T$. A local
        minimum appears as $T$ increases.}
        \label{Fig2}
\end{figure}

Ten quantum wires, labeled (i)--(x) throughout this paper (seven from T622 and
three from T623), were measured in two dilution refrigerators (with base electron
temperature 60 mK and $\sim$150 mK), using standard lock-in techniques and varying
$T$, $B$, $V_{\text{sd}}$, and $n_{\text{\tiny{2D}}}$. Following a mesa etch,
recessed ohmic contacts (Ni/AuGe/Ni/Ti/Pt) were deposited and annealed
\cite{Sarkozy07etal}. A voltage $V_{\text{\tiny{gate}}}$ can be applied to surface
Ti/Au split gates of length $L=400$ nm with width $W=700~(400)$ nm on on T622
(T623). Polyimide insulated the inducing Ti/Au top gate from other gates and ohmic
contacts.

Although the average distance between impurities is $D\!\geq\!0.6\,\mu$m, their
distribution is not uniform. In analogy to mean-free-path calculations, the
probability $P$ of finding an impurity within a 1D channel of length $L$ is
$P=1-e^{(L/D)}\sim$ 50\%. For $G\!\leq\!0.8\,G_0$, an unsplit, symmetric ZBA was
observed in all ten devices. Figure 2(a) shows the ZBA in eight of these. It is
thus unlikely (of order $\prod_{j=1}^{10} P_j\!\ll\!1\%$) that \textit{all} such
occurrences were the result of interactions between conduction electrons and some
localized state near the 1D channel.

\begin{figure}
    \includegraphics[width=\columnwidth]{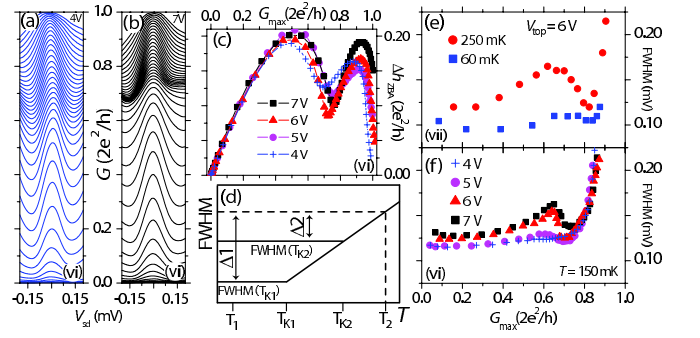}
        \caption{(color online) $G$ vs.~$V_{\text{sd}}$ incrementing
        $V_{\text{\tiny{gate}}}$ ($T\!\approx\!150$\,mK) for: (a)
        $V_{\text{\tiny{top}}}=+4$\,V
        ($n_{\text{\tiny{2D}}}=0.8\times$10$^{11}$\,cm$^{-2}$),
        and (b) $V_{\text{\tiny{top}}}=+7$\,V
        ($n_{\text{\tiny{2D}}}=1.6\times$10$^{11}$\,cm$^{-2}$).
        (c) $\Delta h_{\text{\tiny{ZBA}}}$ for
        $V_{\text{\tiny{top}}}=4-7$\,V. (d) Sketch showing
        \textsc{fwhm}~$\propto \text{max}[T,T_{\text{\tiny{K}}}]$ as $T$ increases.
        (e) \textsc{fwhm} of the ZBA for $T=60$\,mK and 250\,mK. (f) \textsc{fwhm} for
        $V_{\text{\tiny{top}}}=4-7$\,V from the dataset in panel (c).}
        \label{Fig3}
\end{figure}

Defining $G_{\text{\tiny{max}}}$ as the maximum conductance achieved at base $T$,
$V_{\text{sd}}\!=\!0$, and $B\!=\!0$ for each value of $V_{\text{\tiny{gate}}}$,
Fig.~2(b) shows that $G_{\text{\tiny{max}}}$ increases monotonically with
$V_{\!\text{\tiny{gate}}}$ (as in all our devices). Defining $\Delta
h_{\text{\tiny{ZBA}}}$ as $G_{\text{\tiny{max}}}$ minus the average conductance of
the local minima on the \textsc{rhs} and \textsc{lhs} of the ZBA, Fig.~2(d) shows
that $\Delta h_{\text{\tiny{ZBA}}}$ decreases as $T$ increases for all
$V_{\text{\tiny{gate}}}$, as would be expected from Kondo physics. As $T$
increases, a local minimum near $G_{\text{\tiny{max}}}\approx 0.75\,G_0$ becomes
more pronounced. In a previous study on doped quantum wires (see Fig.~6 in
Ref.~\cite{Francois08Aetal}), similar plots of $\Delta h_{\text{\tiny{ZBA}}}$ also
showed a local minimum near $G_{\text{\tiny{max}}}\approx 0.75\,G_0$. Figure 2(c)
links its appearance to the formation of the 0.7 structure.

Varying $n_{\text{\tiny{2D}}}$ affects the Fermi energy of electrons entering the
1D channel from the 2D leads, as well as the 1D confinement potential [e.g.
increasing $V_{\text{\tiny{top}}}=4$\,V in Fig.~3(a) to $7$\,V in Fig.~3(b), the
energy-level spacing between the first two 1D subbands increases from 0.6 to
0.8\,meV]. Figure 3(c) shows no clear trend for $\Delta h_{\text{\tiny{ZBA}}}$ with
increasing $n_{\text{\tiny{2D}}}$, but the minimum near $G_{\text{\tiny{max}}}
\approx 0.75\,G_0$ remains present in all curves. In the Kondo formalism
[Fig.~3(d)], a specific $T_{\text{\tiny{K}}}$ is associated with each
$V_{\text{\tiny{gate}}}$, and the full width at half maximum (\textsc{fwhm}) of the
ZBA should scale linearly either with its $T_{\text{\tiny{K}}}$ if
$T_{\text{\tiny{K}}}\!>\!T$, or with $T$ if $T\!>\!T_{\text{\tiny{K}}}$
\cite{Glazman88-A,Cronenwett98etal}. For $G_{\text{\tiny{max}}}\!\geq\!0.9\,G_0$ in
Fig.~3(f), we do not use the \textsc{fwhm} as it is difficult to distinguish the
ZBA unambiguously from the bell-shape traces of $G$ just below a plateau (see
Fig.~6 in Ref.~\cite{Martin-Moreno92}). For $G_{\text{\tiny{max}}}\!<\!0.7\,G_0$ at
$V_{\text{\tiny{top}}}=4$\,V, the \textsc{fwhm} remain essentially flat:
$T\!>\!T_{\text{\tiny{K}}}$. For $0.5\,G_0\!<\!G_{\text{\tiny{max}}}\!<\!0.7\,G_0$,
increasing $n_{\text{\tiny{2D}}}$ appears to increase $T_{\text{\tiny{K}}}$ beyond
$T\!\approx\!150$\,mK. An upper limit of
$T_{\text{\tiny{K}}}\!<\!\frac{\text{\textsc{fwhm}}}{k_{\text{\tiny{B}}}}$ at each
$V_{\!\text{\tiny{gate}}}$ can be estimated \cite{vanWiel00etal}. In most devices,
regardless of whether the 0.7 structure is visible or not, the \textsc{fwhm} has a
local minimum near $G_{\text{\tiny{max}}} \approx 0.75\,G_0$. Identical minima are
also observed in doped GaAs quantum wires (see Fig.~3 in
Ref.~\cite{Cronenwett02etal}) and in GaN quantum wires (see Fig.~4 in
Ref.~\cite{Chou05Aetal}). Near $G_{\text{\tiny{max}}}\approx 0.75\,G_0$, we
interpret the \textsc{fwhm} minimum to indicate a suppression of Kondo
interactions, leading to a non-monotonic increase of
$T_{\text{\tiny{K}}}(V_{\text{\tiny{gate}}})$ from pinch-off to $2e^2/h$, in direct
contradiction to 1D Kondo theory \cite{Meir02}. Kondo theory also predicts that
\textsc{fwhm}($T_{\text{\tiny{K1}}}$) will increase more than
\textsc{fwhm}($T_{\text{\tiny{K2}}}$) as $T$ increases [i.e. $\Delta 1 > \Delta 2$
in Fig.~3(d)]. However, in further disagreement with theory, Fig.~3(e) shows the
\textit{opposite} behavior: the \textsc{fwhm}s associated with the larger Kondo
temperatures increase the most.

\begin{figure}
    \includegraphics[width=\columnwidth]{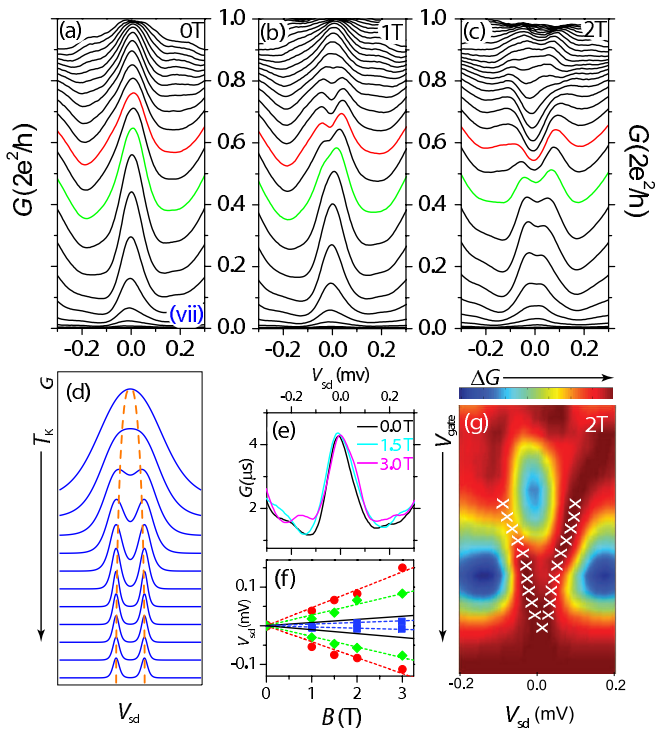}
        \caption{(color) $G$ vs.~$V_{\text{sd}}$ incrementing
        $V_{\text{\tiny{gate}}}$ ($T\!=\!60$\,mK) for:
        (a) $B=0$\,T, (b) $B=1$\,T, and (c) $B=2$\,T.
        (d) Sketch of the expected splitting of the ZBA at constant $B$ and $T$
        for the singlet Kondo effect as $T_{\text{\tiny{K}}}$ alone is decreased
        from top to bottom (traces offset vertically).
        (e) Enlarged view of the ZBA being barely spin-split near pinch-off for
        device (vii).
        (f) Zeeman splitting of the ZBA as a function of $B$ for the red
        and green traces in panels (a)--(c). The black solid line shows the expected
        peak-splitting $g\mu_{\text{\tiny{B}}}/e=25$ $\mu$V/T (for $|g|=0.44$).
        The blue squares are from the data in panel (e).
        (g) Colorscale of the data from panel (c). The ``$\times$'' symbols mark
        the location of the spin-split ZBA peaks.}
        \label{Fig4}
\end{figure}

Figures 4(a)--(c) show how the ZBA spin-splits at low $B$. At a fixed $B$, the
peak-to-peak separation $\Delta V_{\text{p-p}}$ increases almost linearly with
$V_{\text{\tiny{gate}}}$ [Fig.~4(g)]. In an in-plane $B$, pinch-off voltage can
change due to diamagnetic shift \cite{Stern68}, making
$V_{\text{\tiny{gate}}}$ an unreliable marker. However,
$G(|V_{\text{sd}}|>0.25\text{\,mV})$ is mostly insensitive to $B$, while the ZBA
changes significantly. Thus, fitting the linear relation $\Delta
V_{\text{p-p}}=\alpha B$ to the red points in Fig.~4(f), obtained from all red
traces with $G=0.65\,G_0$ at $V_{\text{sd}}=0.25$ mV in Figs.~4(a)--(c), yields
$\alpha=(86\pm2)$ $\mu$V/T. For all green traces with $G=0.50\,G_0$ at
$V_{\text{sd}}=0.25$ mV in Figs.~4(a)--(c), Fig.~4(f) yields
$\alpha=(57\pm2)$ $\mu$V/T. As $V_{\text{\tiny{gate}}}$ decreases [from the red
traces in Figs.~4(a)--(c) down to pinch-off], $\alpha$ appears to continuously
decrease from 86 $\mu$V/T to small values [e.g.~$\alpha < (16\pm5)$ $\mu$V/T from
peak-fitting two asymmetric gaussians to Fig.~4(e)].

\begin{figure}
    \includegraphics[width=\columnwidth]{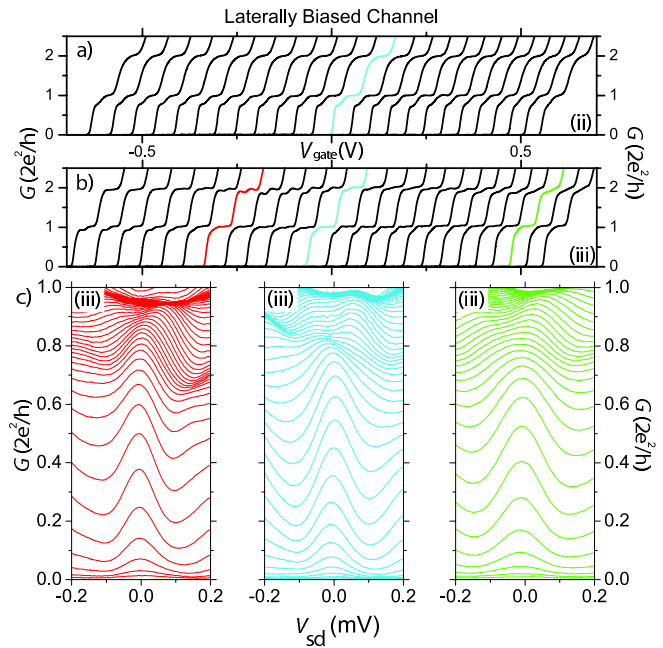}
        \caption{(color) At $T\approx 150$ mK, a clean, ``classic'' 0.7 structure (a) can be
         distinguished from disorder effects (b) by laterally shifting the conducting
         1D channel by differentially biasing the left and right gates by
         $\Delta V_{g}=V_{\text{left}}-V_{\text{right}}$ (traces offset laterally).
         Blue traces in both panels correspond to $\Delta V_{g}=0$, and the leftmost
         (rightmost) trace to $\Delta V_{g}=+1.2$\,V ($-$1.2\,V).
         (c) $G$ vs. $V_{\text{sd}}$ incrementing $V_{\text{\tiny{gate}}}$
         corresponding to the red, blue, and green traces from panel (b). The apparent
         splitting at high $G$ is related to disorder.}
\end{figure}

At finite $B$, the ZBA in quantum dots splits into two peaks
\cite{Cronenwett98etal}, whose peak-to-peak separation $e\Delta
V_{\text{p-p}}=2g^*\mu_{\text{\tiny{B}}} B$ is a defining characteristic of the
Kondo effect \cite{Meir93} where $\mu_{\text{\tiny{B}}}$ is the Bohr magneton and
$g^*$ the effective Land\'e $g$ factor. Figure 4(d) illustrates three distinct
regimes one would expect from the singlet Kondo effect at fixed $B$ and $T$
\cite{Pustilnik04-A,Potok07etal}. In the topmost traces,
$k_{\text{\tiny{B}}}T_{\text{\tiny{K}}} > g^*\mu_{\text{\tiny{B}}} B >
k_{\text{\tiny{B}}}T$: spin-splitting cannot be resolved. In the middle traces,
$g^*\mu_{\text{\tiny{B}}} B > k_{\text{\tiny{B}}}T_{\text{\tiny{K}}} >
k_{\text{\tiny{B}}}T$: the linewidth of each split peak is narrow enough to make
the splitting visible. In the bottom traces,
$g^*\mu_{\text{\tiny{B}}}B>k_{\text{\tiny{B}}}T>
k_{\text{\tiny{B}}}T_{\text{\tiny{K}}}$: the split peaks shrink \textit{but their
splitting should remain constant} as long they are still resolvable. However, in
our quantum wires, this is clearly not the case. The variation of $\Delta
V_{\text{p-p}}=\alpha B$ with $V_{\text{\tiny{gate}}}$ in Fig.~4(b)--(c) cannot be
reconciled with singlet Kondo physics.

In quantum dots, the ZBA splitting can vary with $V_{\text{\tiny{gate}}}$ for $B
\geq 0$ (Fig.~4 in Ref.~\cite{Jeong01}, Fig.~3 in Ref.~\cite{Chen04}) from the
competition between the Kondo effect and the Ruderman-Kittel-Kasuya-Yosida (RKKY)
interaction between two localized spins \cite{DiasDaSilva06}. Although two such
localised spins are predicted to form in quantum wires near pinch-off
\cite{Rejec06,Berggren08} and these could explain the behavior observed in
Figs.~4(b)--(c), this scenario would also require the ZBA to be split at $B=0$.
This is not the case [Figs.~2(b), 3(a)--(b), 4(a), 5(c)]: the two-impurity Kondo model
is not applicable.

In spin-polarization models
\cite{Wang96,Kristensen00etal,Reilly02etal,Abi07,Francois08Betal,Berggren08}, the
energy difference between spin-up and spin-down electrons $\Delta
E_{\uparrow\downarrow} = g\mu_{\text{\tiny{B}}}B +
E_{\text{ex}}(n_{\text{\tiny{1D}}})$ includes $E_{\text{ex}}$, an exchange-enhanced
spin splitting that could account for previous observations of an enhanced $g$
factor above the value $|g|=0.44$ of bulk GaAs \cite{Thomas96etal}. Neglecting
correlation effects, the bare exchange energy in 1D scales linearly with
$n_{\text{\tiny{1D}}}$. Assuming $n_{\text{\tiny{1D}}} \propto
V_{\text{\tiny{gate}}}$, the almost linear splitting of the ZBA is consistent with
a density-dependent spin polarization. However, this scenario would also require
that the minimum value of $e\alpha$ be the bare Zeeman energy
$g\mu_{\text{\tiny{B}}} = 25$ $\mu\,$eV/T. This is not what we observe: $e\alpha
<16$ $\mu\,$eV/T in Fig.~4(e). Instead, we find $\Delta E_{\uparrow\downarrow} =
g^*(n_{\text{\tiny{1D}}})\mu_{\text{\tiny{B}}}B$, where
$0.27<g^*(n_{\text{\tiny{1D}}})<1.5$ [Fig.~4(f)]. The Zeeman effect can be
suppressed ($g^*\sim 0.2$) if a 2DEG significantly penetrates into the AlGaAs
barriers \cite{Kogan04etal}, at high $n_{\text{\tiny{2D}}}$ or if the 2DEG is close
to the surface. Neither situation applies to our devices. The suppression of the
bare Zeeman effect at pinch-off in our quantum wires is not consistent with spin
polarization models.

Despite their exceptional device-to-device reproducibility (compared with doped
wires), undoped quantum wires are not free from disorder [Fig.~5(b)]. The apparent
splitting for $G\!\geq\!0.8\,G_0$ in some of our devices [Fig.~5(c)] is not due to
spontaneous spin-splitting or RKKY vs.~Kondo interactions, but rather to resonant
backscattering or length resonances \cite{Lindelof08}. By increasing the 2D density
(and thus long-range screening), many disorder-related effects can be minimized.

In summary, we provide compelling evidence for the ZBA to be a fundamental property
of quantum wires. Its continued presence from $G \sim 2e^2/h$ down to $G \sim
(2e^2/h)\times 10^{-5}$ suggests it is a different phenomenon to the 0.7 structure,
as proposed in \cite{Yoon07etal,Francois08Aetal}. Both 1D Kondo physics and spin
polarization models fall short of accurately predicting experimental observations.
For 1D Kondo physics models, these are: (i) a non-monotonic increase of
$T_{\text{\tiny{K}}}$ with $V_{\text{\tiny{gate}}}$, (ii) the \textsc{fwhm} of the ZBA
not scaling with max[$T,T_{\text{\tiny{K}}}$] as $T$ increases, and (iii) a linear
peak-splitting of the ZBA with $V_{\text{\tiny{gate}}}$ at fixed $B$. Spin
polarization models can account neither for the occurrence of the ZBA nor for the
suppression of the bare Zeeman effect at pinch-off. It is hoped that further
refinements in theory will account for these observations.

The authors acknowledge D. Anderson, H. Quach and C. Namba for electron beam
patterning, and V. Tripathi, K.-F. Berggren, A.R. Hamilton, C.J.B. Ford, J.P.
Griffiths, T.M. Chen, K.J. Thomas, and N.R. Cooper for useful discussions. S. Sarkozy
acknowledges financial support as a Northrop Grumman Space Technology Doctoral
Fellow. I. Farrer thanks Toshiba Research Europe for financial support.

\bibliographystyle{apsrev}
%\bibliography{../refmaster}

\begin{thebibliography}{37}
\expandafter\ifx\csname natexlab\endcsname\relax\def\natexlab#1{#1}\fi
\expandafter\ifx\csname bibnamefont\endcsname\relax
  \def\bibnamefont#1{#1}\fi
\expandafter\ifx\csname bibfnamefont\endcsname\relax
  \def\bibfnamefont#1{#1}\fi
\expandafter\ifx\csname citenamefont\endcsname\relax
  \def\citenamefont#1{#1}\fi
\expandafter\ifx\csname url\endcsname\relax
  \def\url#1{\texttt{#1}}\fi
\expandafter\ifx\csname urlprefix\endcsname\relax\def\urlprefix{URL }\fi
\providecommand{\bibinfo}[2]{#2} \providecommand{\eprint}[2][]{\url{#2}}

\bibitem[{\citenamefont{Thornton et~al.}(1986)\citenamefont{Thornton, Pepper,
  Ahmed, Andrews, and Davies}}]{Thornton86}
\bibinfo{author}{\bibfnamefont{T.~J.} \bibnamefont{Thornton}},
  \bibinfo{author}{\bibfnamefont{M.}~\bibnamefont{Pepper}},
  \bibinfo{author}{\bibfnamefont{H.}~\bibnamefont{Ahmed}},
  \bibinfo{author}{\bibfnamefont{D.}~\bibnamefont{Andrews}}, \bibnamefont{and}
  \bibinfo{author}{\bibfnamefont{G.~J.} \bibnamefont{Davies}},
  \bibinfo{journal}{Phys. Rev. Lett.} \textbf{\bibinfo{volume}{56}},
  \bibinfo{pages}{1198} (\bibinfo{year}{1986}).

\bibitem[{\citenamefont{van Wees{\textit{ et al.}}}(1988)}]{vanWees88etal}
    \bibinfo{author}{\bibfnamefont{B.~J.} \bibnamefont{van Wees{\textit{ et
  al.}}}}, \bibinfo{journal}{Phys. Rev. Lett.} \textbf{\bibinfo{volume}{60}},
  \bibinfo{pages}{848} (\bibinfo{year}{1988}).

\bibitem[{\citenamefont{Wharam{\textit{ et al.}}}(1988)}]{Wharam88etal}
    \bibinfo{author}{\bibfnamefont{D.~A.} \bibnamefont{Wharam{\textit{ et al.}}}},
  \bibinfo{journal}{J. Phys. C} \textbf{\bibinfo{volume}{21}},
  \bibinfo{pages}{L209} (\bibinfo{year}{1988}).

\bibitem[{\citenamefont{Thomas{\textit{ et al.}}}(1996)}]{Thomas96etal}
    \bibinfo{author}{\bibfnamefont{K.~J.} \bibnamefont{Thomas{\textit{ et al.}}}},
  \bibinfo{journal}{Phys. Rev. Lett.} \textbf{\bibinfo{volume}{77}},
  \bibinfo{pages}{135} (\bibinfo{year}{1996}).

\bibitem[{\citenamefont{Wang and Berggren}(1996)}]{Wang96}
    \bibinfo{author}{\bibfnamefont{C.~K.} \bibnamefont{Wang}} \bibnamefont{and}
  \bibinfo{author}{\bibfnamefont{K.~F.} \bibnamefont{Berggren}},
  \bibinfo{journal}{Phys. Rev. B} \textbf{\bibinfo{volume}{54}},
  \bibinfo{pages}{R14257} (\bibinfo{year}{1996}).

\bibitem[{\citenamefont{Kristensen{\textit{ et al.}}}(2000)}]{Kristensen00etal}
    \bibinfo{author}{\bibfnamefont{A.}~\bibnamefont{Kristensen{\textit{ et al.}}}},
  \bibinfo{journal}{Phys. Rev. B} \textbf{\bibinfo{volume}{62}},
  \bibinfo{pages}{10950} (\bibinfo{year}{2000}).

\bibitem[{\citenamefont{Reilly{\textit{ et al.}}}(2002)}]{Reilly02etal}
    \bibinfo{author}{\bibfnamefont{D.~J.} \bibnamefont{Reilly{\textit{ et al.}}}},
  \bibinfo{journal}{Phys. Rev. Lett.} \textbf{\bibinfo{volume}{89}},
  \bibinfo{pages}{246801} (\bibinfo{year}{2002}).

\bibitem[{\citenamefont{Graham et~al.}(2007)\citenamefont{Graham, Sawkey,
  Pepper, Simmons, and Ritchie}}]{Abi07}
\bibinfo{author}{\bibfnamefont{A.~C.} \bibnamefont{Graham}},
  \bibinfo{author}{\bibfnamefont{D.~L.} \bibnamefont{Sawkey}},
  \bibinfo{author}{\bibfnamefont{M.}~\bibnamefont{Pepper}},
  \bibinfo{author}{\bibfnamefont{M.~Y.} \bibnamefont{Simmons}},
  \bibnamefont{and} \bibinfo{author}{\bibfnamefont{D.~A.}
  \bibnamefont{Ritchie}}, \bibinfo{journal}{Phys. Rev. B}
  \textbf{\bibinfo{volume}{75}}, \bibinfo{pages}{035331}
  (\bibinfo{year}{2007}).

\bibitem[{\citenamefont{Sfigakis{\textit{ et
  al.}}}(2008{\natexlab{a}})}]{Francois08Betal}
\bibinfo{author}{\bibfnamefont{F.}~\bibnamefont{Sfigakis{\textit{ et al.}}}},
  \bibinfo{journal}{J. Phys.: Condens. Matter} \textbf{\bibinfo{volume}{20}},
  \bibinfo{pages}{164213} (\bibinfo{year}{2008}{\natexlab{a}}).

\bibitem[{\citenamefont{Berggren and Yakimenko}(2008)}]{Berggren08}
    \bibinfo{author}{\bibfnamefont{K.~F.} \bibnamefont{Berggren}} \bibnamefont{and}
  \bibinfo{author}{\bibfnamefont{I.}~\bibnamefont{Yakimenko}},
  \bibinfo{journal}{J. Phys.: Condens. Matter} \textbf{\bibinfo{volume}{20}},
  \bibinfo{pages}{164203} (\bibinfo{year}{2008}).

\bibitem[{\citenamefont{Cronenwett{\textit{ et al.}}}(2002)}]{Cronenwett02etal}
    \bibinfo{author}{\bibfnamefont{S.}~\bibnamefont{Cronenwett{\textit{ et al.}}}},
  \bibinfo{journal}{Phys. Rev. Lett.} \textbf{\bibinfo{volume}{88}},
  \bibinfo{pages}{226805} (\bibinfo{year}{2002}).

\bibitem[{\citenamefont{Meir et~al.}(2002)\citenamefont{Meir, Hirose, and
  Wingreen}}]{Meir02}
\bibinfo{author}{\bibfnamefont{Y.}~\bibnamefont{Meir}},
  \bibinfo{author}{\bibfnamefont{K.}~\bibnamefont{Hirose}}, \bibnamefont{and}
  \bibinfo{author}{\bibfnamefont{N.~S.} \bibnamefont{Wingreen}},
  \bibinfo{journal}{Phys. Rev. Lett.} \textbf{\bibinfo{volume}{89}},
  \bibinfo{pages}{196802} (\bibinfo{year}{2002}).

\bibitem[{\citenamefont{Rejec and Meir}(2006)}]{Rejec06}
    \bibinfo{author}{\bibfnamefont{T.}~\bibnamefont{Rejec}} \bibnamefont{and}
  \bibinfo{author}{\bibfnamefont{Y.}~\bibnamefont{Meir}},
  \bibinfo{journal}{Nature} \textbf{\bibinfo{volume}{442}},
  \bibinfo{pages}{900} (\bibinfo{year}{2006}).

\bibitem[{\citenamefont{Meir et~al.}(1993)\citenamefont{Meir, Wingreen, and
  Lee}}]{Meir93}
\bibinfo{author}{\bibfnamefont{Y.}~\bibnamefont{Meir}},
  \bibinfo{author}{\bibfnamefont{N.~S.} \bibnamefont{Wingreen}},
  \bibnamefont{and} \bibinfo{author}{\bibfnamefont{P.~A.} \bibnamefont{Lee}},
  \bibinfo{journal}{Phys. Rev. Lett.} \textbf{\bibinfo{volume}{70}},
  \bibinfo{pages}{2601} (\bibinfo{year}{1993}).

\bibitem[{\citenamefont{Goldhaber-Gordon{\textit{ et
  al.}}}(1998)}]{Goldhaber98Aetal}
\bibinfo{author}{\bibfnamefont{D.}~\bibnamefont{Goldhaber-Gordon{\textit{ et
  al.}}}}, \bibinfo{journal}{Nature} \textbf{\bibinfo{volume}{391}},
  \bibinfo{pages}{156} (\bibinfo{year}{1998}).

\bibitem[{\citenamefont{Cronenwett{\textit{ et al.}}}(1998)}]{Cronenwett98etal}
    \bibinfo{author}{\bibfnamefont{S.~M.} \bibnamefont{Cronenwett{\textit{ et
  al.}}}}, \bibinfo{journal}{Science} \textbf{\bibinfo{volume}{281}},
  \bibinfo{pages}{540} (\bibinfo{year}{1998}).

\bibitem[{\citenamefont{van~der Wiel{\textit{ et al.}}}(2000)}]{vanWiel00etal}
    \bibinfo{author}{\bibfnamefont{W.~G.} \bibnamefont{van~der Wiel{\textit{ et
  al.}}}}, \bibinfo{journal}{Science} \textbf{\bibinfo{volume}{289}},
  \bibinfo{pages}{2105} (\bibinfo{year}{2000}).

\bibitem[{\citenamefont{Yoon{ \textit{et al.}}}(2007)}]{Yoon07etal}
    \bibinfo{author}{\bibfnamefont{Y.}~\bibnamefont{Yoon{ \textit{et al.}}}},
  \bibinfo{journal}{Phys. Rev. Lett.} \textbf{\bibinfo{volume}{99}},
  \bibinfo{pages}{136805} (\bibinfo{year}{2007}).

\bibitem[{\citenamefont{Sfigakis{\textit{ et
  al.}}}(2008{\natexlab{b}})}]{Francois08Aetal}
\bibinfo{author}{\bibfnamefont{F.}~\bibnamefont{Sfigakis{\textit{ et al.}}}},
  \bibinfo{journal}{Phys. Rev. Lett.} \textbf{\bibinfo{volume}{100}},
  \bibinfo{pages}{026807} (\bibinfo{year}{2008}{\natexlab{b}}).

\bibitem[{\citenamefont{Graham{\textit{ et al.}}}(2008)}]{AbiPRBsubmit}
    \bibinfo{author}{\bibfnamefont{A.~C.} \bibnamefont{Graham{\textit{ et al.}}}},
  \bibinfo{journal}{submitted to Phys. Rev. B}  (\bibinfo{year}{2008}).

\bibitem[{\citenamefont{Griffiths}(unpublished)}]{JonGnopaper}
    \bibinfo{author}{\bibfnamefont{J.~P.} \bibnamefont{Griffiths}}
  (\bibinfo{year}{unpublished}).

\bibitem[{\citenamefont{Harrell{\textit{ et al.}}}(1999)}]{Harrell99etal}
    \bibinfo{author}{\bibfnamefont{R.~H.} \bibnamefont{Harrell{\textit{ et al.}}}},
  \bibinfo{journal}{Appl. Phys. Lett.} \textbf{\bibinfo{volume}{74}},
  \bibinfo{pages}{2328} (\bibinfo{year}{1999}).

\bibitem[{\citenamefont{Sarkozy{\textit{ et al.}}}(2008)}]{SarkozyAPLsubmit}
    \bibinfo{author}{\bibfnamefont{S.}~\bibnamefont{Sarkozy{\textit{ et al.}}}},
  \bibinfo{journal}{submitted to Appl. Phys. Lett.}  (\bibinfo{year}{2008}).

\bibitem[{\citenamefont{Ando et~al.}(1982)\citenamefont{Ando, Fowler, and
  Stern}}]{Ando82-A}
\bibinfo{author}{\bibfnamefont{T.}~\bibnamefont{Ando}},
  \bibinfo{author}{\bibfnamefont{A.~B.} \bibnamefont{Fowler}},
  \bibnamefont{and} \bibinfo{author}{\bibfnamefont{F.}~\bibnamefont{Stern}},
  \bibinfo{journal}{Rev. Mod. Phys.} \textbf{\bibinfo{volume}{54}},
  \bibinfo{pages}{437} (\bibinfo{year}{1982}).

\bibitem[{\citenamefont{Gold}(1988)}]{Gold88}
    \bibinfo{author}{\bibfnamefont{A.}~\bibnamefont{Gold}}, \bibinfo{journal}{Phys.
  Rev. B} \textbf{\bibinfo{volume}{38}}, \bibinfo{pages}{10798}
  (\bibinfo{year}{1988}).

\bibitem[{\citenamefont{Sarkozy{\textit{ et al}}}(2007)}]{Sarkozy07etal}
    \bibinfo{author}{\bibfnamefont{S.}~\bibnamefont{Sarkozy{\textit{ et al}}}},
  \bibinfo{journal}{Electrochemical Soc Proc.} \textbf{\bibinfo{volume}{11}},
  \bibinfo{pages}{75} (\bibinfo{year}{2007}).

\bibitem[{\citenamefont{Glazman and Raikh}(1988)}]{Glazman88-A}
    \bibinfo{author}{\bibfnamefont{L.}~\bibnamefont{Glazman}} \bibnamefont{and}
  \bibinfo{author}{\bibfnamefont{M.}~\bibnamefont{Raikh}},
  \bibinfo{journal}{JETP Letters} \textbf{\bibinfo{volume}{47}},
  \bibinfo{pages}{452} (\bibinfo{year}{1988}).

\bibitem[{\citenamefont{Martin-Moreno et~al.}(1992)\citenamefont{Martin-Moreno,
  Nicholls, Patel, and Pepper}}]{Martin-Moreno92}
\bibinfo{author}{\bibfnamefont{L.}~\bibnamefont{Martin-Moreno}},
  \bibinfo{author}{\bibfnamefont{J.~T.} \bibnamefont{Nicholls}},
  \bibinfo{author}{\bibfnamefont{N.~K.} \bibnamefont{Patel}}, \bibnamefont{and}
  \bibinfo{author}{\bibfnamefont{M.}~\bibnamefont{Pepper}},
  \bibinfo{journal}{J. Phys.: Condens. Matter} \textbf{\bibinfo{volume}{4}},
  \bibinfo{pages}{1323} (\bibinfo{year}{1992}).

\bibitem[{\citenamefont{Chou{\textit{ et al.}}}(2005)}]{Chou05Aetal}
    \bibinfo{author}{\bibfnamefont{H.~T.} \bibnamefont{Chou{\textit{ et al.}}}},
  \bibinfo{journal}{Appl. Phys. Lett.} \textbf{\bibinfo{volume}{86}},
  \bibinfo{pages}{073108} (\bibinfo{year}{2005}).

\bibitem[{\citenamefont{Stern}(1968)}]{Stern68}
    \bibinfo{author}{\bibfnamefont{F.}~\bibnamefont{Stern}},
  \bibinfo{journal}{Phys. Rev. Lett.} \textbf{\bibinfo{volume}{21}},
  \bibinfo{pages}{1687} (\bibinfo{year}{1968}).

\bibitem[{\citenamefont{Pustilnik and Glazman}(2004)}]{Pustilnik04-A}
    \bibinfo{author}{\bibfnamefont{M.}~\bibnamefont{Pustilnik}} \bibnamefont{and}
  \bibinfo{author}{\bibfnamefont{L.~I.} \bibnamefont{Glazman}},
  \bibinfo{journal}{J. Phys.: Condens. Matter} \textbf{\bibinfo{volume}{16}},
  \bibinfo{pages}{R513} (\bibinfo{year}{2004}).

\bibitem[{\citenamefont{Potok{\textit{ et al.}}}(2007)}]{Potok07etal}
    \bibinfo{author}{\bibfnamefont{R.~M.} \bibnamefont{Potok{\textit{ et al.}}}},
  \bibinfo{journal}{Nature} \textbf{\bibinfo{volume}{446}},
  \bibinfo{pages}{167} (\bibinfo{year}{2007}).

\bibitem[{\citenamefont{Jeong et~al.}(2001)\citenamefont{Jeong, Chang, and
  Melloch}}]{Jeong01}
\bibinfo{author}{\bibfnamefont{H.}~\bibnamefont{Jeong}},
  \bibinfo{author}{\bibfnamefont{A.~M.} \bibnamefont{Chang}}, \bibnamefont{and}
  \bibinfo{author}{\bibfnamefont{M.~R.} \bibnamefont{Melloch}},
  \bibinfo{journal}{Science} \textbf{\bibinfo{volume}{293}},
  \bibinfo{pages}{2221} (\bibinfo{year}{2001}).

\bibitem[{\citenamefont{Chen et~al.}(2004)\citenamefont{Chen, Chang, and
  Melloch}}]{Chen04}
\bibinfo{author}{\bibfnamefont{J.~C.} \bibnamefont{Chen}},
  \bibinfo{author}{\bibfnamefont{A.~M.} \bibnamefont{Chang}}, \bibnamefont{and}
  \bibinfo{author}{\bibfnamefont{M.~R.} \bibnamefont{Melloch}},
  \bibinfo{journal}{Phys. Rev. Lett.} \textbf{\bibinfo{volume}{92}},
  \bibinfo{pages}{176801} (\bibinfo{year}{2004}).

\bibitem[{\citenamefont{{Dias~da~Silva}
  et~al.}(2006)\citenamefont{{Dias~da~Silva}, Sandler, Ingersent, and
  Ulloa}}]{DiasDaSilva06}
\bibinfo{author}{\bibfnamefont{L.~G. G.~V.} \bibnamefont{{Dias~da~Silva}}},
  \bibinfo{author}{\bibfnamefont{N.~P.} \bibnamefont{Sandler}},
  \bibinfo{author}{\bibfnamefont{K.}~\bibnamefont{Ingersent}},
  \bibnamefont{and} \bibinfo{author}{\bibfnamefont{S.~E.} \bibnamefont{Ulloa}},
  \bibinfo{journal}{Phys. Rev. Lett.} \textbf{\bibinfo{volume}{97}},
  \bibinfo{pages}{096603} (\bibinfo{year}{2006}).

\bibitem[{\citenamefont{Kogan{\textit{ et al.}}}(2004)}]{Kogan04etal}
    \bibinfo{author}{\bibfnamefont{A.}~\bibnamefont{Kogan{\textit{ et al.}}}},
  \bibinfo{journal}{Phys. Rev. Lett.} \textbf{\bibinfo{volume}{93}},
  \bibinfo{pages}{166602} (\bibinfo{year}{2004}).

\bibitem[{\citenamefont{Lindelof and Aagesen}(2008)}]{Lindelof08}
    \bibinfo{author}{\bibfnamefont{P.~E.} \bibnamefont{Lindelof}} \bibnamefont{and}
  \bibinfo{author}{\bibfnamefont{M.}~\bibnamefont{Aagesen}},
  \bibinfo{journal}{J. Phys.: Condens. Matter} \textbf{\bibinfo{volume}{20}},
  \bibinfo{pages}{164207} (\bibinfo{year}{2008}).

\end{thebibliography}

\end{document}